\documentclass[letterpaper,twocolumn,showpacs,amsmath,pra,amssymb]{revtex4-1} 
\usepackage[T1]{fontenc}
\usepackage[latin9]{inputenc} 
\usepackage{times}
\usepackage{color}
\usepackage{xspace}
\usepackage{amssymb,amsmath}
\usepackage{amsbsy}
\usepackage{graphicx}
\usepackage{epstopdf}
\usepackage{float}

\newcommand{\diff}{\mathrm{d}}

\newcommand{\bx}{\ensuremath{\mathbf{x}}\xspace}

\renewcommand{\k}{\mathbf{k}}

\begin{document}
\newcommand{\bk}{\mathbf{k}}
\newcommand{\nt}{\tilde{n}}
\newcommand{\Lnh}{\ensuremath{\mathcal{L}}\xspace}
\newcommand{\dbx}{\diff\bx}
\newcommand{\dbk}{\diff\bk}
\newcommand{\nbe}{\ensuremath{\bar{n}_{\mathrm{BE}}}\xspace}

\title{Thermally induced coherence in a Mott insulator of bosonic atoms}
\author{E.~Toth} 
\author{P.~B.~Blakie}  
\affiliation{Jack Dodd Centre for Quantum Technology, Department of Physics, University of Otago, Dunedin, New Zealand.}

\begin{abstract}
Conventional wisdom is that increasing temperature causes quantum coherence to decrease.
 Using finite temperature perturbation theory and exact calculations for the strongly correlated bosonic Mott insulating state  we show a practical counter-example that can be explored in optical lattice experiments:  the short-range coherence of the Mott insulating phase can increase substantially with increasing temperature. 
We demonstrate that  this phenomenon originates from thermally produced defects that can tunnel with ease. Since the near zero temperature coherence properties have been measured with high precision we expect these results to be  verifiable in current experiments.
\end{abstract}
\maketitle

\section{Introduction}
\begin{figure}[!t]
\begin{center}
\includegraphics[width=2.9in]{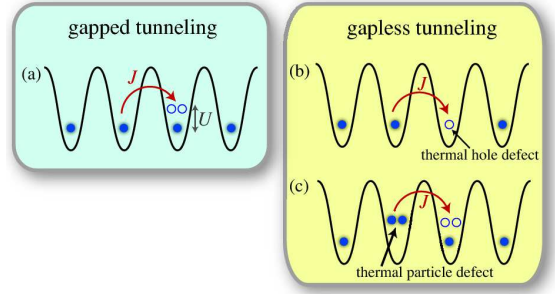}
\caption{(Color online) Schematic of tunneling processes that lead to short-range coherence being established. (a) \textit{Gapped tunneling}: Particle-hole formation caused by  tunneling, $J$, but costing the gap energy $U$. (b,c) \textit{Gapless tunneling}: Thermally activated defects  can tunnel  without energy cost.}
\label{fig:therm_latt}
\end{center}
\end{figure}

A key signature of the superfluid to Mott insulator phase transition was the disappearance of long-range phase coherence, identified through the loss of sharp peaks in the interference pattern  after expansion \cite{Greiner2002a}.  Subsequent experiments showed \cite{Gerbier2005a,Spielman2007a} that a weak and slowly modulated inference pattern remained in the insulating regime, arising from short-range coherence between neighboring sites established by virtual hopping events [see Fig.~\ref{fig:therm_latt}(a)]. These processes occur with amplitude $J/U$, where $J$ and $U$ are the tunneling matrix element and onsite interaction energy, respectively \cite{Jaksch1998a,*Blakie2004a}.  Since $J/U$  is small in the insulating regime, this tunneling process is well described by first order perturbation theory.
Experiments in two-dimensional (2D) \cite{Spielman2007a} and three-dimensional (3D) \cite{Gerbier2005a} lattices have accurately measured this coherence and obtained quantitative agreement with the predicted $J/U$ scaling.

Another topic of much recent interest in optical lattice research has been the influence of temperature on the properties  and phases of the many body system \cite{Rigol2005a,*Buonsante2007a,*Diener2007a,*Koponen2007a,*Baillie2009a}, with considerable development of techniques for measuring thermal properties \cite{Weld2009a,*McKay2009a}.  
An area of particular interest has been the use of lattice loading \cite{Blakie2004b,*Ho2007a,*Pollet2008a,*Gerbier2007b} or other methods \cite{Ho2009a,*Weld2010a} to produce states of low temperature/entropy, e.g.~to produce a system suitable for simulating spin Hamiltonians \cite{Sansone2010a}.

Here we study the influence of temperature on short-range coherence in the Mott insulating state. The physics we explore is a competition between two processes: (i) \textit{Gapped tunneling}, whereby an energy cost of $U$ is needed for a particle to tunnel to its neighbor [Fig.~\ref{fig:therm_latt}(a)]. As mentioned above, this occurs with amplitude $J/U$ and is the dominant process at $T=0$. (ii) \textit{Gapless tunneling}, occurs when the unperturbed state has a hole [Fig.~\ref{fig:therm_latt}(b)] or particle [Fig.~\ref{fig:therm_latt}(c)] defect which can tunnel  with no energy cost and contribute strongly to coherence. However, the production of such defects are exponentially suppressed at low temperatures. Here we study the Mott insulator coherence using finite temperature perturbation theory and exact solutions for small systems. We find that  
at $T\sim0.1U/k_B$, gapless tunneling becomes important and can lead to an appreciable increase in the short-range coherence  over the zero temperature value. 
We present results for both homogeneous (translationally invariant) and inhomogenous (harmonically trapped) lattices like in experiments \cite{Spielman2007a}.
 \section{Theory}
 We first consider a system of bosons in a transitionally invariant lattice described by the Bose-Hubbard Hamiltonian
 \begin{equation}
 \hat{H}=-J\sum_{\langle i j\rangle}\hat{a}^\dagger_i\hat{a}_j+\frac{U}{2}\sum_j\hat{a}^\dagger_j\hat{a}^\dagger_j\hat{a}_j\hat{a}_j.
 \end{equation}
 We assume for simplicity that the lattice is simple cubic in geometry and note that the tunneling is limited to nearest neighbor lattice sites. 
Our regime of interest is the Mott insulating phase ($J/U\ll1$) with an integer number ($\bar{n}$) of atoms per site. To describe the short-range coherence between nearest neighbor sites, we evaluate the correlation function 
 \begin{equation}
 c_1\equiv \langle \hat{a}^\dagger_i\hat{a}_{i^\prime}\rangle,
 \end{equation}
  where ${i^\prime}$ is taken to be a  nearest neighbor lattice site of $i$.  
\subsection{ Zero temperature} We first review  the zero temperature limit (see \cite{Gerbier2005a,Spielman2007a,Gerbier2005b,Toth2008a}). In this case, to lowest order in $J$, the manybody state is a \textit{perfect} Mott insulator with precisely $\bar{n}$ atoms per site, i.e.~
$|\psi^{(0)}\rangle=\prod_j \left[(\hat{a}_j^\dagger )^{\bar{n}}/{\sqrt{\bar{n}!}}\right]|\rm{vac}\rangle$,
 for which $c_1^{(0)}=0$. First order corrections to the ground state arise from the tunneling of a particle between neighboring sites. This process is gapped (costing energy  $U$ to create a particle-hole excitation) and so the amplitude for the particle-hole admixture [see Fig.~\ref{fig:therm_latt}(a)] is $\sqrt{\bar{n}(\bar{n}+1)}J/U$ (according first order perturbation theory), giving $c^{(1)}_1={2\bar{n}(\bar{n}+1)}J/U$.
 
 The effect of this neighboring site coherence is directly observable in the momentum distribution as
 \begin{equation}
 \rho(\mathbf{k})=N|w(\k)|^2\left[1+\frac{2c_1}{c_0}\sum_{\alpha=x,y,z}\cos(k_\alpha a)\right],\label{homogrho}
 \end{equation}
where $N$ is the total number of atoms, $c_0=\langle \hat{a}_i^\dagger\hat{a}_i\rangle$ (which is $\bar{n}$ in the Mott insulating regime), $w(\k)$ is the Fourier transform of the Wannier state (the localized state at each lattice site), and $a$ is the lattice constant. By fitting the momentum distribution $c_1$ was extracted in Ref.~\cite{Spielman2007a}  verifying the zero temperature prediction. Equivalently    $c_1$  can be determined from the visibility of the interference fringes (e.g.~see \cite{Gerbier2005a,Gerbier2005b,Hoffmann2009a}).

Higher-order processes can establish coherence to next-nearest neighbors, leading to more rapid modulations of the momentum distribution [e.g.~$\cos(2k_\alpha a)$]. However, these contributions are much smaller and thus more difficult to measure except for larger values of $J/U$ where the system is near the superfluid transition. For this reason we  focus upon the nearest neighbor coherence $c_1$.
\subsection{Finite temperature} 
 At finite temperature the first order correlation function is  
 \begin{equation}
 c_1=\langle \hat{a}_i^\dagger\hat{a}_{i'}\rangle=\mathcal{Z}^{-1}\mathrm{Tr}\{\hat{a}_i^\dagger\hat{a}_{i'}e^{-\beta(\hat{H}-\mu\hat{N})}\},\label{finiteTcorrl}
 \end{equation}
 where $\mathcal{Z}=\mathrm{Tr}\{e^{-\beta(\hat{H}-\mu\hat{N})}\}$ is the partition function, with $\beta=1/k_BT$, $\hat{N}=\sum_j\hat{a}^\dagger_j\hat{a}_j$, and $\mu$ the chemical potential, used to adjust the average number of particles.
 
\noindent {\bf Neglecting tunneling:}  
By neglecting tunneling we obtain a system which can be solved exactly, but has no short-range coherence $c_1=0$. However, this system is useful to consider as it is the unperturbed system for  a perturbative expansion in the tunneling parameter.

For $J=0$ the lattice sites decouple, and the problem reduces to a single site anharmonic oscillator with energy levels
\begin{equation} 
E_n  =\frac{1}{2}Un(n-1)-\mu n, 
\end{equation}
for which the partition function is  
 $Z^{(0)} =\sum_{n}e^{-\beta E_n}.$
The mean number of atoms per site is given by $c_0=\sum_nn\rho_n$, where $\rho_{n}=e^{-\beta E_n}/Z^{(0)}$ is the Gibbs factor. 
 \begin{figure}[!t]
\begin{center}
\includegraphics[width=4in]{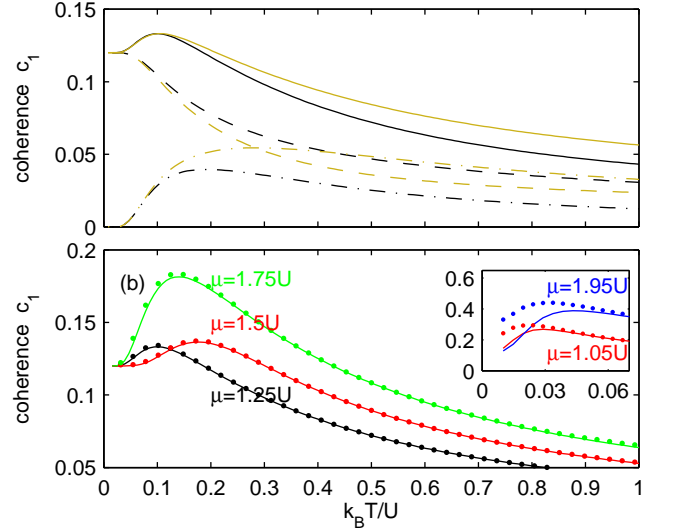}
\caption{(Color online) Nearest neighbor coherence. (a) Perturbative treatment: (solid) full result $c_1^{(1)}$, (dashed) gapped  $c_1^{(1a)}$, and (dash-dot) gapless  $c_1^{(1b)}$ contributions for the case of $\mu=1.5U$. 
(Black) show full perturbative calculation [Eqs.~(\ref{c11a2}), (\ref{c11b2})] while (yellow/grey) show the simple analytic results  [Eqs.~(\ref{c11aA}), (\ref{c11bA})]. (b) Variation with chemical potential and comparison of (solid) full perturbative results against (dots) exact calculations for a 1D lattice. $\mu=1.25U$ (black), $1.5U$  (red/grey), and $1.75U$  (green/light grey). {Inset: comparison of (solid) full perturbative results against (dots) exact calculations near the phase boundary  $\mu=1.95U$ (blue/gray), $1.05U$  (red/light grey).}
Other parameters: $J=0.01U$ and $\bar{n}=2$. 
}
\label{fig:coher}
\end{center}
\end{figure}
 
\noindent {\bf First order perturbative treatment:} 
We use imaginary time perturbation theory   to obtain an expansion for (\ref{finiteTcorrl}). At first order in $J$ we have
 \begin{equation}
 c_1^{(1)}= J\sum_{n n'}n'(n+1)\rho_{n}\rho_{n'}\int_0^{\beta}d\tau\, e^{-\tau(\epsilon_{n}^+ +\epsilon_{n'}^{-} )},\label{c11}
 \end{equation}
 where $n$ ($n'$) is the number of atoms at site $i$ ($i'$),   and  we have introduced the  particle and hole energies given by $\epsilon^+_n \equiv E_{n+1}-E_n,$ and $\epsilon^-_n \equiv E_{n-1}-E_n,$ respectively. Note that $\epsilon_{n}^+ +\epsilon_{n'}^{-} $ is the energy required (at zeroth order in $J$) to tunnel a particle from site $i'$ to $i$, i.e.~effect the change $(n,n')\to(n+1,n'-1)$.  Eq.~(\ref{c11}) is derived in a similar manner to expression (18) in \cite{Hoffmann2009a} and (44) in \cite{Freericks2009a}, and we do not repeat the details here.
 
 To evaluate the integral in Eq.~(\ref{c11}) we note the \textit{energy cost} term is given by  $\epsilon_{n}^+ +\epsilon_{n'}^{-} =U(1+n-n')$, and thus two cases emerge: (a) Gapped excitations with $n\ne n'-1$, and (b) gapless excitations for the case $n=n'-1$. 
The first order coherence arising from the gapped excitations is given by
\begin{align}
c_1^{(1a)} 
 =& \frac{2J}{U}\sum_{n n'}\frac{n'(n+1)\rho_{n}\rho_{n'}}{1+n-n'} (1-\delta_{n,n'-1}),\label{c11a2}
 \end{align}
while for the gapless excitations   \begin{align}
 c_1^{(1b)}  =& \beta J\sum_{n}(n+1)^2\rho_{n}\rho_{n+1}.\label{c11b2}
\end{align}
In Fig.~\ref{fig:coher}(a) we show $c_1$ and the separate contributions of the gapped and gapless parts, while in Fig.~\ref{fig:coher}(b) we examine the dependence of $c_1$ on chemical potential. The key result of the paper, which we further analyze below, is that the gapless contribution at finite temperature serves to enhance system coherence over the $T=0$ state. For the various values of chemical potential we find that the maximum temperature induced coherence occurs for $T\approx [0.1 - 0.2]U/k_{B}$, while the melting temperature for a trapped Mott insulator is  $\sim0.2U/k_{B}$  \cite{Gerbier2007b}.

We note that, by not explicitly accounting for the gapless excitations, the first order coherence expressions developed in \cite{Hoffmann2009a} and \cite{Freericks2009a} are formally divergent in the finite temperature regime. Indeed, results for the finite temperature visibility presented in Fig.~4 of \cite{Hoffmann2009a} do not show the enhancement we observe at finite temperature, and instead show the characteristic behavior of the gapless  part $(c^{(1a)})$ only.

Since the first order expression for the coherence involves only a single tunneling event, it is independent of lattice dimension. Higher order processes can explore the various possible paths to tunnel between lattice sites, which is strongly dependent on the lattice dimension and geometry.  However, the next correction to $c_1$ is at 3rd order and for low temperatures and small $J/U$ this should be negligible.

\noindent {\bf Simple analytic treatment:}  
 Here we present an analytic simplification  of results (\ref{c11a2}) and (\ref{c11b2}) that helps clarify the underlying physics. In the low temperature limit,  $k_BT\ll U$, the summation in $c_1^{(1a)}$ is dominated by the $n=n'=\bar{n}$ term (since $\rho_{\bar{n}}\approx 1$) and we have
 \begin{equation}
 c_1^{(1a)} \approx \frac{2J}{U} \bar{n}(\bar{n}+1)\rho_{\bar{n}}^2, \label{c11aA}
 \end{equation}
  which reduces to the known zero temperature result as $T\to0$. The summation in $c_1^{(1b)}$ is dominated by the $n=n'\pm1=\bar{n}$ terms, for which $\rho_{\bar{n}\pm1}\approx e^{-\beta \epsilon^{\pm}_{\bar{n}}}$ and we have
 \begin{equation}
 c_1^{(1b)} \approx \beta J  \rho_{\bar{n}}\{(\bar{n}+1)^2\rho_{\bar{n}+1}+\bar{n}^2\rho_{\bar{n}-1}\}. \label{c11bA}
 \end{equation}
Due to the  energetic degeneracy  of the gapless states coupled, this term occurs with a prefactor of $\beta J$  rather than  $J/U$, however because defect states (which allow gapless tunneling) are thermally generated, the are exponentially suppressed by the factor $\rho_{\bar{n}\pm1}$ with temperature, and hence  $c_1^{(1b)}\to0$ as $T\to0$. 

In Fig.~\ref{fig:coher}(a) we compare the simplified analytic results to the perturbative calculations. To do this we have taken $\rho_{\bar{n}}=1/z$, $\rho_{\bar{n}\pm1}=e^{-\beta\epsilon^{\pm}_{\bar{n}}}/z$, where $z=1+e^{-\beta\epsilon_{\bar{n}}^+}+e^{-\beta\epsilon_{\bar{n}}^-}$. Agreement is quantitatively good up until $T\sim0.2U/k_B$, as above this temperature the $\rho_{\bar{n}\pm1}$ terms are becoming comparable to $\rho_{\bar{n}}$, and other terms neglected  in our simplified treatment become important as the system begins to  melt.

The $\bar{n}$ atom-per-site Mott state occurs for chemical potentials in the range $\mu/U\in(\bar{n}-1,\bar{n})$. The results in Fig.~\ref{fig:coher}(a) are evaluated in the middle of the $\bar{n}=2$ range. It is of interest to understand how the coherence varies with chemical potential for application to the trapped case (see Sec.~\ref{sec:Incoh}). In Fig.~\ref{fig:coher}(b) we consider the behavior of $c_1^{(1)}$ for several values of chemical potential. These results show that the chemical potential can considerably enhance the finite temperature coherence peak. This is easily understood as changing the chemical potential makes it easier to produce defects by altering the values of $\epsilon^{\pm}_{\bar{n}}$,
e.g., for the case of $\mu=1.75U$ we have that $\epsilon^{+}_{\bar{n}}=0.25U$ (c.f. $\epsilon^{-}_{\bar{n}}=0.75U$), leading to an  enhancement of the first term in Eq.~(\ref{c11bA}). 
We note that the perturbative expansion is not valid when $\mu/U$ is nearly integer, and in the trapped system so-called \textit{superfluid shells} form where the local chemical potential satisfies this condition.

\noindent {\bf Nonperturbative  treatment:} 
We also solve Eq.~(\ref{finiteTcorrl}) by exact diagonalization for small 1D systems in a lattice with $M$ sites and periodic boundary conditions.
 In practice we do this using the truncated Fock basis  $|n_1,\ldots,n_M\rangle$ with $0\le n_j\le n_{\max}$ to  represent the grand canonical Hamiltonian which we then numerically diagonalize. We retain the $N_E$ lowest energy eigenvalues and vectors, $\{\mathcal{E}_\alpha,|\psi_\alpha\rangle\}$, and  construct the density matrix
 $\rho=\tilde{\mathcal{Z}}^{-1}\sum_{\alpha=1}^{N_E}e^{-\beta \mathcal{E}_\alpha}|\psi_\alpha\rangle\langle\psi_\alpha|,$
 with $\tilde{\mathcal{Z}}=\sum_{\alpha=1}^{N_E}e^{-\beta \mathcal{E}_\alpha}$, from which we calculate  $c_1$.
 There are two important criteria for the validity of this calculation to describe $c_{1}$ in the uniform (infinite) 1D system: (i) 
 $M\gg1$, so that finite size effects are negligible for the calculation of $c_1$ \cite{Damski2006a}; (ii) The basis is sufficiently large to include all energy scales accessed at finite temperature, i.e.~$[Un_{\max}(n_{\max}-1)/2,\mathcal{E}_{N_E}-\mathcal{E}_1]\gg k_BT$.

 In Fig.~\ref{fig:coher}(b) we compare the perturbation theory against the analytic results using $M=5$, $n_{\max}=7$, and $N_E=620$ for which $\mathcal{E}_{N_{E}}-\mathcal{E}_1\ge 5U$.
 We find good agreement between the exact and perturbation results up to the maximum temperatures considered, with the differences always remaining less than 3\%, demonstrating that for these values of tunneling the system is well within the perturbative regime.  
  We find that perturbation theory starts to breakdown for $J/U\gtrsim0.1$. 
The inset of Fig.~\ref{fig:coher}(b) considers cases where $\mu/U$  approaches an integer value. This shows that perturbation theory breaks down near the phase boundary for low temperature, and care must be taken to exclude such regions in the extension of the perturbation theory to a inhomogeneous system \cite{ Gerbier2005b}.
 While our comparisons here are for a one-dimensional (1D) lattice, it is worth noting that the agreement between perturbative and exact treatments is expected to improve with increasing lattice dimension  \cite{Freericks2009a}.

\subsection{Inhomogeneous system}\label{sec:Incoh}
\begin{figure}[!t]
\begin{center}
\includegraphics[width=3.7in]{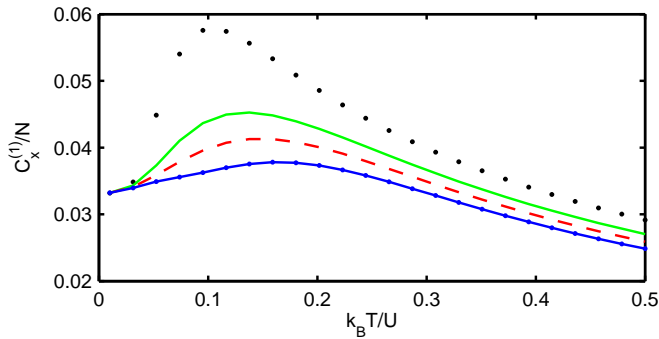}
\caption{(Color online) Nearest neighbor coherence in the $x$ direction.
(Black dots) Homogeneous case [independent of dimension] with $\mu=0.8U$.
Inhomogeneous 1D case for $\mu=0.8U$  (green/light gray line), and for $\mu=0.5U$ (blue/black dotted line).  Inhomogeneous 2D case for
  $\mu=0.8U$  (red/grey dashed line).  Other parameters: $J/U=8.35\times10^{-3}$, $\mu$ values chosen so that $\bar{n}=1$ at trap center, interaction parameters are for $^{87}$Rb in a lattice produced using counter-propagating $\lambda=820$ nm lasers, and 
harmonic confinement frequency is $\omega=2\pi \times40$ Hz. 
 }
\label{fig:Incoh}
\end{center}
\end{figure}

In experiment an additional external potential $V(\mathbf{r})$  is present due to magnetic confinement and/or the Gaussian profile of the lattice beams. This can be represented in our theory by the replacement $\mu\hat{N}\to\sum_j\mu_j\hat{a}^\dagger_j\hat{a}_j$ in (\ref{finiteTcorrl}), where $\mu_j=\mu-V(\mathbf{r}_j)$ is the local chemical potential, with $V(\mathbf{r}_j)$ the value of the external potential at lattice site $j$.  The generalization of Eq.~(\ref{homogrho}) is then
  \begin{equation}
 \rho(\mathbf{k})=N|w(\k)|^2\left[1+\sum_{\alpha}\frac{2C_{1,\alpha}}{N}\cos(k_\alpha a)\right],
 \end{equation}
where $C_{1,\alpha}=\sum_{\langle ii'\rangle_\alpha}\langle \hat{a}_{i}^\dagger\hat{a}_{i'}\rangle$, with ${\langle ii'\rangle_\alpha}$ indicating a sum over nearest neighbor lattice sites along the $\alpha$  direction.
In experiments \cite{Greiner2002a, Gerbier2005a, Spielman2007a} the external potential varies slowly across the system, which allows us to use the local density approximation. Thus we have $C_{1,\alpha}=2\int^{\mu}_{-\infty} d\mu_{r}g(\mu_{r})c_{1}(\mu_{r})$, where $g(\mu_{r})$ is density of trap-potential states. We present results for $C_{1,x}^{(1)}/N$ (calculated using first order perturbation theory\footnote{In the inhomogeneous calculations we exclude the small superfluid shells (where perturbation theory breaks down) and only include the $\bar{n}=1$ Mott insulator region (similar to Ref.~\cite{Gerbier2005b}). 
}) in various systems in Fig.~\ref{fig:Incoh},
where we notice a decrease in the peak value for the inhomogeneous case.  This can be explained by the tendency of the spatially varying chemical potential to smear out the temperature induced coherence [e.g.~consider the curves for various $\mu$ values in Fig.~\ref{fig:coher}(b)]. This smearing is least pronounced for a 1D system as the density of states is strongly peaked around the chemical potential at the trap center [in contrast to 2D where the $g(\mu_{r})$ is evenly distributed across $\mu_{r}$]. From Fig.~\ref{fig:Incoh} we can see that the enhancement of short-range coherence is around  $\sim15\%$ of the $T=0$ value  for the 2D case and  $\sim28\%$ in the 1D system. This could be easily observable with current experiments, particularly with the recent experimental improvements reported in \cite{Garcia2010a}.

\section{Conclusions}
In this paper we have studied the role of temperature on the coherence of a strongly correlated quantum system. Our main tool, finite temperature perturbation theory, provides insight into the underlying physical origin of the increase in coherence at finite temperature: gapless tunneling of thermally activated defects.  We have verified our predictions against exact numerical calculations.
While the formalism for finite temperature perturbation theory has been developed by several other authors, it is usually applied at $T=0$ with gapped excitations. The only application to finite temperature regime (that we are aware of) was a calculation for $c_1$  in Ref.~\cite{Hoffmann2009a}. Their results disagree with our work (predicting $c_1$  to decrease with $T$)  due to neglect  of gapless excitations.

Our study shows that the equilibrium state coherence of a Mott insulator state increases with temperature, proportional to the number of thermally induced defects. 
Interestingly, if defects are produced in the system by other means e.g.~through non-adiabatic changes  of the lattice potential, or three-body loss, then additional enhancement of the coherence is expected.  Thus coherence measurements might complement \textit{in situ} imaging \cite{Bakr2009a,*Sherson2010a} to characterize defects in experiments.

\noindent {\bf Acknowledgements:} 
  This work was supported by Marsden contract 09-UOO-093, and FRST contract NERF-UOOX0703.

\end{document}